\documentclass[aps,prb,twocolumn,superscriptaddress,longbibliography]{revtex4-2}
\usepackage[colorlinks=true,linkcolor=blue,citecolor=blue]{hyperref}
\usepackage{amsmath,amssymb}
\usepackage[utf8]{inputenc}
\usepackage[T1]{fontenc}
\usepackage{url}
\usepackage{graphicx}
\usepackage{bm}
\usepackage{bbm}
\usepackage{xcolor}
\usepackage{color}
\usepackage{ulem}

\begin{document}

\title{Topological triple phase transition in non-Hermitian quasicrystals with complex asymmetric hopping}

\author{Shaina Gandhi}
\email{p20200058@pilani.bits-pilani.ac.in}
\affiliation{Department of Physics, Birla Institute of Technology and Science, Pilani 333031, India} 

\author{Jayendra N. Bandyopadhyay}
\email{jnbandyo@gmail.com}
\affiliation{Department of Physics, Birla Institute of Technology and Science, Pilani 333031, India}

\begin{abstract}
 
The triple phase transitions or simultaneous transitions of three different phases, namely topological, parity-time (PT) symmetry breaking, and metal-insulator transitions, are observed in an extension of PT symmetric non-Hermitian Aubry-Andr\'e-Harper model. In this model, besides non-Hermitian complex quasi-periodic onsite potential, non-Hermiticity is also included in the nearest-neighbor hopping terms. Moreover, the nearest-neighbor hopping terms is also quasi-periodic. The presence of two non-Hermitian parameters, one from the onsite potential and another one from the hopping part, ensures PT symmetry transition in the system. In addition, tuning these two non-Hermitian parameters, we identify a parameters regime, where we observe the triple phase transition. Following some recent studies, an electrical circuit based experimental realization of this model is also discussed.

\end{abstract}

\maketitle 

In quantum mechanics, the Hermitian property of any observable not only ensures the measurement outcomes of that observable to be real, but it also preserves the total probability of all the outcomes. Over the past few years, systems with non-Hermitian Hamiltonians have become an important field of research \cite{shen2018topological,bergholtz2021exceptional,ghatak2019new,kawabata2019symmetry}. The non-Hermitian Hamiltonians effectively represent quantum systems which exchange particles and/or energy with their environment  \cite{ashida2020non, bergholtz2021exceptional}. The non-Hermiticity in the Hamiltonians leads to the complex energy spectra which is a signature of non-conservative systems. However, the non-Hermitian Hamiltonians with parity-time (PT) symmetry have real eigenvalues \cite{bender1998real, el2018non, bender2002generalized}. Therefore, the PT symmetric Hamiltonians are considered as a more general class of Hamiltonians which allows loss and gain in the systems.

The spectral properties of the non-Hermitian systems show some strange behaviors under the presence of different symmetries, which has no Hermitian counterpart. One prominent example is that non-Hermitian Hamiltonians with PT symmetry host a typical special degeneracy, known as exceptional points (EPs), where the eigenvalues and the eigenvectors coalesce and thus the corresponding non-Hermitian matrices do not have a full basis of eigenstates \cite{wang2021topological, li2020symmetry, yuce2018pt, jin2017schrieffer, zhu2014pt, xu2020fate}. Furthermore, the skin effect in non-Hermitian systems emerges from their inherent topology. This phenomenon refers to the localization of eigenstates near the system's boundary. This localization is a manifestation of the nontrivial topological properties of the system, which can be characterized using mathematical tools such as topological invariants \cite{PhysRevLett.124.086801, PhysRevLett.125.126402}. Specifically, under open boundary conditions, the bulk states concentrate at the edges of the lattice, giving rise to the non-Hermitian skin effect. This phenomenon breaks the conventional bulk-boundary correspondence (BBC) observed in Hermitian systems \cite{yao2018edge, song2019non, lee2016anomalous}. 

Previous investigations in the field of non-Hermitian quantum mechanics have provided valuable insights into the intriguing phenomena that arise when quantum mechanical particles encounter randomness and non-Hermitian systems. A notable and extensively studied example in this realm is the non-Hermitian version of the celebrated Anderson model \cite{PhysRev.109.1492}. The non-Hermitian Anderson model has also been studied extensively \cite{PhysRevLett.77.570, PhysRevB.56.8651, HATANO1998317, PhysRevB.56.R4333,PhysRevB.101.014202, PhysRevB.101.014204, PhysRevLett.126.166801, PhysRevLett.126.090402}. Notably, Hatano and Nelson were pioneers in exploring the localization transitions in quantum mechanical particles described by the non-Hermitian Anderson model with asymmetric hopping \cite{PhysRevLett.77.570, PhysRevB.56.8651,HATANO1998317}. Their groundbreaking study revealed a localization-delocalization transition induced by the presence of the random potential, characterized by the emergence of complex eigenvalues in the system.
Subsequently, the investigation of the non-Hermitian Anderson model expanded to include two-dimensional  \cite{PhysRevB.101.014202} and three-dimensional cases \cite{PhysRevB.101.014204, PhysRevLett.126.090402}, providing further insights into the localization properties of the eigenstates. These models, unlike the Hatano-Nelson model, incorporated non-Hermiticity by introducing complex onsite random potentials.

In addition, there is another class of non-Hermitian models in 1D, known as Aubry-Andr\'e-Harper (NH-AAH) models \cite{yuce2014pt, zeng2017anderson, zhang2020non, jiang2019interplay, tang2021localization}.
The NH-AAH model physically represents 1D quasicrystals (QCs) with loss and gain processes. The quasicrystalline property and the non-Hermiticity are introduced into the system through an onsite quasi-periodic potential by a complex phase factor. The Hermitian counterpart of this model shows metal-insulator (MI) transition determined by a fixed ratio of the strength of the onsite quasi-periodic potential and the nearest-neighbor hopping strength \cite{aubry1980analyticity}. Besides this ratio, the MI transition in the NH-AAH model with PT symmetry also depends on the imaginary part of the complex phase. Very interestingly, a recent study has found that the NH-AAH model preserves PT symmetry when the system is in the metallic phase. This implies that at the MI phase transition point, the PT symmetry in the NH-AAH model spontaneously breaks down via exceptional points and complex energy eigenvalues start appearing \cite{longhi2019metal, PhysRevLett.77.570, PhysRevB.56.8651}. Moreover, it is worth noting that at the dual transition point where both the MI phase transition and PT symmetry breaking occur, the NH-AAH model exhibits an additional topological phase transition \cite{PhysRevLett.122.237601}. This combination of three simultaneous phase transitions, often referred to as the ``{\it triple}'' phase transition, has attracted significant interest in the field. Furthermore, experimental realizations of this triple phase transition have been reported \cite{weidemann2022topological}, providing further validation of the theoretical predictions. These three phase transitions have completely different underlying physics: (a) The MI transition is determined by the strength of disorder in the system; (b) the PT symmetry breaking leads to real to complex energy spectrum, and (c) the topological phase transition is determined by the loss of adiabatic connection of the Hamiltonian at the transition point. The non-Hermiticity is responsible for this rare event in the NH-AAH Hamiltonian.

Recently, a non-Hermitian version of the AAH model with PT symmetry is proposed \cite{PhysRevA.103.L011302}. In this model, the non-Hermiticity is introduced via asymmetric complex hopping. This model shows concurrent transitions of MI and PT symmetry. These phase transitions are robust against the system's size and the onsite disorder represented by incommensurate (quasi-periodic) potential.
In Ref. \cite{PhysRevB.104.024201}, another version of the non-Hermitian AA model was proposed. In this model, the non-Hermiticity was introduced at onsite quasi-periodic potential, as well as at the non-reciprocal hopping. Here, the interplay of the non-Hermiticity and the disorder captures a new physical aspect. For any values of the system's parameters, this model is not PT symmetric. The presence of two non-Hermitian parameters is responsible for this non-PT symmetric nature of the model. As a consequence, the triple phase transition cannot be observed in this model.
 
The aim of this paper is to propose a NH-AAH model with PT symmetry while incorporating two non-Hermitian parameters. The two non-Hermitian parameters are introduced in two ways: the standard onsite quasi-periodic potential with a complex phase factor, and an incommensurately modulated complex asymmetric hopping. Remarkably, this system exhibits a triple phase transition within a specific regime of the system's parameters. 

The tight-binding Hamiltonian of this NH-AAH model is
\begin{equation}
 \begin{split}
 H &= \sum_{j=1}^L V \cos(2\pi \beta j + \phi_1 +\phi_2) \hat{c}_j^{\dagger} \hat{c}_j +\\ &+ [t+i\gamma \sin(2\pi \beta j + \phi_1)]  \hat{c}_{j+1}^{\dagger} \hat{c}_j
 \\ &+ [t+i\gamma \sin(2\pi \beta (j+1) + \phi_1)]  \hat{c}_j^{\dagger} \hat{c}_{j+1},
  \end{split}
  \label{1}
\end{equation}  
where the phase factors are $\phi_1 = \pi - \pi \beta \,\,({\rm mod}\, 2\pi)$ and $\phi_2 = \theta - ih$. The number of lattice sites is $L$. With this choice of the parameter $\phi_1$ \cite{PhysRevA.103.L011302} and setting $\theta = 0$, the above Hamiltonian becomes PT symmetric. The space-reflection operator $\hat{P}$ and the time-reversal operator $\hat{T}$ have the following effects: $\hat{P}^{-1} \hat{c}_j \hat{P} = \hat{c}_{L+1-j}$, and $\hat{T}^{-1} i\, \hat{T} = -i$. We have provided a detailed analytical demonstration of the PT symmetry of our model in Appendix \ref{appendixA}. Here, the parameters $h$ and $\gamma$ respectively determine the strength of the non-Hermiticity at the onsite potential and hopping. Here, the onsite potential is modulated by a cosine function of strength $V$ with periodicity $\frac{1}{\beta}$. The second and third terms of the Hamiltonian describe the complex asymmetric hopping.   

We chose to set the parameter $\beta$ in our study to be equal to $\sigma_G$, which is the inverse of the golden mean. This choice of $\beta$ is a popular and commonly used option in research on the AAH model. The golden mean, as well as its inverse $\sigma_G$, satisfy diophantine condition. The best diophantine (rational) approximation of $\sigma_G$ is the ratio of two consecutive numbers of the Fibonacci series $\{f_n\}$ and the approximation becomes better with increasing $n$, i.e., $\left|\sigma_G -\lim\limits_{n\rightarrow\infty} (f_{n-1}/f_n)\right| \rightarrow 0$. The Fibonacci series is obtained from the recursion relation $f_n = f_{n-1} + f_{n-2}$, with initial values $f_0 = 0$ and $f_1 = 1$. The above diophantine approximation of $\sigma_G$ allows us to consider the following standard practice for the numerics. We set $\beta = f_{n-1}/f_n$ with a sufficiently large $n$ and assume the number of lattice sites $L = f_n$. This approximation not only ensures the quasi-periodicity of the onsite potential for a finite number of lattice sites, but it also makes the dual transformation to the Fourier transformed space exact \cite{Aulbach_2004}. Therefore, in our study, we set $\beta = \frac{144}{233}$, where $144$ and $233$ are consecutive Fibonacci numbers, and fix the number of lattice sites $L=233$ throughout this study.

The Hermitian version of our model (i.e., when $\gamma = h = 0$) with irrational $\beta$  shows the MI transition at the critical point $V=2t$. For $V<2t$, the system is metallic, whereas for $V>2t$ the system becomes an insulator. In the case of the single parameter non-Hermitian case with $\gamma = 0$ and $h \neq 0$, the system is a self-dual. This self-duality is manifested by the eigenfunctions having identical distributions in both real and momentum spaces, indicating an effective mapping between the Hamiltonian and its Fourier transform. Exploiting this self-duality, the critical point of the metal-insulator (MI) transition was analytically calculated as $h = \ln \left(\frac{2t}{V}\right)$ for the NH-AAH model with a complex onsite potential \cite{PhysRevLett.122.237601}. As mentioned earlier that this critical point is the triple phase transition point, i.e., the simultaneous transition point for the MI transition, the PT transition, and the topological phase transitions. Unlike the NH-AAH model, where the non-Hermiticity is introduced solely through a complex onsite potential, our model described by the Hamiltonian given in Eq. (\ref{1}) incorporates both complex sinusoidal asymmetric hopping terms and complex onsite potential. As a result, non-Hermiticity arises in two distinct parameters. Due to this additional complexity, our model does not exhibit self-duality. Hence, analytically we can not find the transition point for this model. Therefore, the phase transition and other properties of this model are studied numerically. Furthermore, unlike Hatano-Nelson model with non-Hermiticity in the off-diagonal disorder \cite{PhysRevLett.77.570, PhysRevB.56.8651} and the NH-AAH model with non-Hermitian onsite quasi-periodic potential  \cite{weidemann2022topological}, our model has non-Hermiticity in both onsite potential and in nearest-neighbor hopping.

\begin{figure}[t]
\centering
\includegraphics[width=8.5cm]{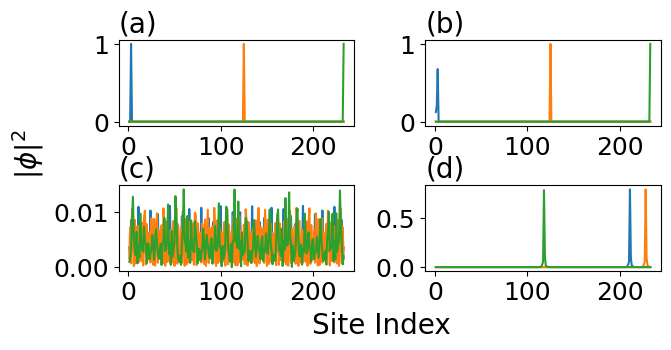}
\caption{Typical nature of the eigenstates corresponding to the Hamiltonian given in Eq. (\ref{1}) with the OBC (top) and the PBC (bottom) are compared for the number of lattice sites $L=233$. (a)-(b) The top row shows three randomly picked eigenstates under the OBC which are all localized at some lattice site for both $h=0.4$ and $h=1.2$, respectively. (c)-(d) The same eigenstates are presented in the bottom figure with the PBC, where all the eigenstates are extended for $h=0.4$, but are localized at different lattice sites for $h = 1.2$.}
\label{mode_profiles} 
\end{figure}

The prime focus of this paper is to explore the existence of triple phase transition in the presence of two non-Hermitian parameters. The eigenstates of this system is presented in Fig. \ref{mode_profiles}. The eigenstates have revealed interesting insights about the behavior of the NH-AAH model. Here, we set the parameters $\gamma=0.05$, $t=1.0$, and $V=1.0$. Specifically, we observe that the eigenstates of the system are primarily localized when open boundary condition (OBC) is imposed. This is illustrated for $h=0.4$ in  Fig. \ref{mode_profiles}(a), and for $h=1.2$ in Fig. \ref{mode_profiles}(b). However, for the periodic boundary condition (PBC), we observe the effect of MI transition in the system. Here, in the metallic regime when the parameter $h=0.4$, we observe extended eigenstates as shown Fig. \ref{mode_profiles}(c). On the other hand, in Fig. \ref{mode_profiles}(d), we show that in the insulating regime with $h=1.2$, all the eigenstates are localized. 
These extreme boundary-condition-sensitive behaviors of the eigenstates are attributed to the interplay of the disorders due to the onsite quasi-periodic potential and the presence of the non-Hermiticity in the system. 
 
\begin{figure}[b]
\hspace{-0.4cm}
    \includegraphics[width=8cm, height=7cm]{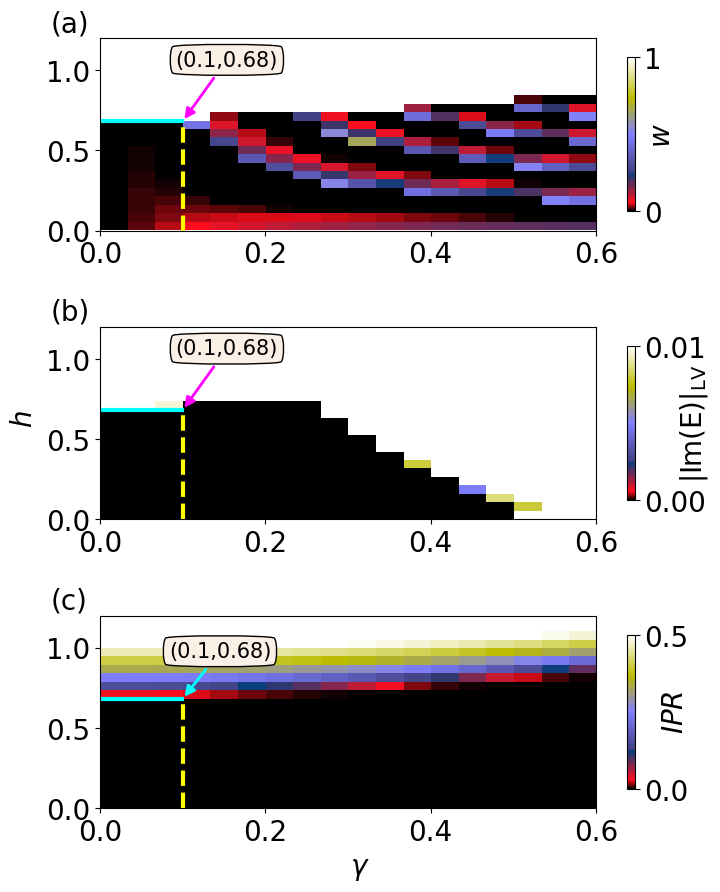}
    \caption{The phase diagrams are presented for the topological, PT symmetry, and the MI transitions as the function of the parameters $\gamma$ and $h$. (a) The topological phases are characterized by the calculation of the winding number $w$ given in Eq. \eqref{wn}. The colors from black to white represent $w=0$ to $1$ values. The fluctuations in the winding number are less for $\gamma<0.1$ (refer color bar). In this region, the phase transition occurs at $h_c = 0.68$ as marked in the figure. (b) The largest value of imaginary parts of the energy eigenvalues are presented to observe the PT symmetry transition. The different color palettes separate real and complex region of the energies (refer color bar). The PT symmetry is preserved in the regime $\gamma<0.5$. A common transition point is observed at $h_c = 0.68$ for $\gamma<0.1$. (c) The maximum value of the inverse participation ratio are shown to track the MI transitions. The MI transitions are observed for the whole range of $\gamma$, but at different values of the parameter $h$. For $\gamma<0.3$, the MI transition occurs at $h_c = 0.68$. The vertical yellow line in all the figures are describing the parameter regime $\gamma <0.1$, where the triple phase transition is possible to observe. The horizontal cyan colored line at $h=h_c=0.68$ marks the triple phase transition points.}
    \label{Phase}   
\end{figure}

\begin{figure}[t]
\begin{tabular}{ccc}
     \hspace*{-0.25cm}
    \includegraphics[width=41mm, height = 3cm]{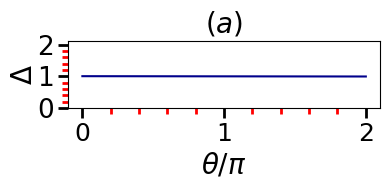}&
	
    \hspace*{-0.28cm}
    \includegraphics[width=40mm, height = 3cm]{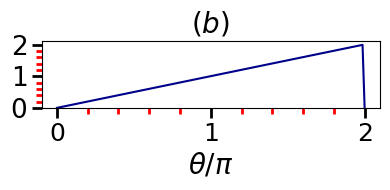} \\
\end{tabular}
\caption{The phase argument of $\rm{det}(H-E_B)$ versus $\theta$ is shown for $h=0.6$ (left) and $h=0.76$ (right) for $\gamma = 0.05$ and the number of lattice sites $L=233$. As $\theta$ increases from $0$ to $2\pi$, the spectral trajectory does not encircles the base energy (left). Effectively, the spectrum of $H$ does not wind. However, for $h=0.76$, the spectral trajectory encircles the base energy once and the corresponding winding number is $1$    }
\label{WnReIm}   
\end{figure}
First we study the variation of the transition point with the parameters $\gamma$ and $h$, while the parameters $V$ and $t$ are fixed at unity, i.e. $V = t = 1.0$. We tune $\gamma$ and $h$ to identify whether the triple phase transition is possible in our model. These results are shown in Fig. \ref{Phase}. In Fig. \ref{Phase}(a), the topological phase transition is presented using winding number as the topological invariant. The winding number is calculated from the following relation \cite{ashida2020non, PhysRevX.8.031079, PhysRevLett.122.237601}:
\begin{equation}
w(h, \gamma) = \lim_{L \to \infty}\frac{1}{2\pi i} \int_{0}^{2\pi} d\theta \frac{\partial}{\partial \theta}\ln \left[\det {H\left(\frac{\theta}{L},h, \gamma\right)- E_B}\right].
\label{wn}
\end{equation}
The winding number $w(h,\gamma)$ counts the number of times the complex spectral trajectory encircles the base energy $E_B$, when the real phase $\theta$ varies from $0$ to $2\pi$. Here, the Hamiltonian $H$ with the PBC is considered and $L$ is as usual the number of lattice sites. For our analysis, we have chosen $E_B = 0$ as the base energy. The strengths of the non-Hermitian parameters decide the values of the winding number. 
We observe smooth topological phase transition identified by the transition of the winding number $0 \leftrightarrow 1$ in the parameter region $\gamma<0.1$ marked by the vertical dashed-yellow line. For the non-Hermitian parameter $\gamma>0.1$, fluctuations in the winding numbers are observed in the topologically trivial region. However, topologically non-trivial region is still very well-protected from the fluctuation in this parameter regime. In the parameter regime $\gamma<0.1$, we find the critical point for the topological phase transition also at $h_c=0.68$, which is shown by the horizontal solid cyan line. 

This behavior of the winding number is further validated in Fig. \ref{WnReIm}, where we set the parameter $\gamma=0.05$. Following Ref. \cite{PhysRevX.8.031079}, in this figure, the argument of determinant of the Hamiltonian $\Delta \equiv {\rm arg} [{\rm det}(H-E_B)]$ with respect to $E_B$ is plotted as a function of the cyclic parameter $\theta$ while traversing $\theta$ for a full cycle of $0$ to $2\pi$. Here, we have selected two values of the parameter $h$: one is $h=0.6$, less than the critical value $h_c$; and the other value is $h=0.76$, larger than $h_c$. For $h=0.6$, presented in Fig. \ref{WnReIm}(a), the winding number $w=0$. Here, we observe that the trajectory of $\Delta$ does not enclose $E_B$ while $\theta$ traverses a complete loop. On the other hand, for $h=0.76$ where $w=1$, the trajectory of $\Delta$ encircles $E_B$ once as shown in Fig. \ref{WnReIm}(b).

Besides topological transition, we observe a simultaneous transition of the PT symmetry breaking as well as the MI phase transition. In Fig. \ref{Phase}(b), we show the PT symmetry transition. Here, we consider the largest value of the imaginary part of energy $|{\rm Im}(E)|_{\rm LV}$ as an indicator of the PT symmetry and study it as a function of the parameters $\gamma$ and $h$. We find that, only in the regime $\gamma<0.5$ (shown in black), the system has pure real eigenenergies. However, here the transition point varies with $\gamma$. Furthermore, we observe that, in $\gamma < 0.1$ regime, the PT symmetry transition point is also at $h_c=0.68$.

In Fig. \ref{Phase}(c), the MI transition is studied in the system investigating localization to delocalization transition of the eigenstates. We use the maximum inverse participation ratio (IPR) of the eigenstates as the measure of localization-delocalization transition. The IPR of a state is defined as
\begin{equation}
{\rm IPR} = \sum_{n}|\psi_n|^4/\Bigl(\sum_{n}(|\psi_n|^2\Bigr)^2.
\label{IPR}
\end{equation}
For the localized case, the IPR is larger and close to {\it unity}. On the other hand, for the delocalized or extended states, the IPR is proportional to the inverse of the dimension, where the dimension is the number of lattice sites $L=233$. Therefore, for the extended or delocalized states (i.e., in the metallic regime), the IPR is very small. In the phase diagram, the metallic phase is represented by the black region and the (strongly) insulating phase region is represented by the white region. The other colors represent the critical regime. 
Here, we observe that the MI transition can occur for the whole range of $\gamma$, but at different values of $h_c$ values. However, again for $\gamma<0.1$, the transition point is at $h_c = 0.68$.

We infer from the phase diagrams of Fig. \ref{Phase} that, in the regime $\gamma <0.1$ and at $h_c=0.68$, the system shows the triple phase transition. Thus, here we observe that, even in the presence of two non-Hermitian parameters, the system can show triple phase transitions. Based on these results, besides $V = t = 1.0$, we fix the parameter $\gamma = 0.05$ and continuously tune the parameter $h$ to monitor the topological, PT symmetric, and MI transitions.

\begin{widetext}
\begin{figure*}[t]        
\begin{tabular}{ccc}
\hspace*{-1cm}
    \includegraphics[width=12cm,height=4cm]{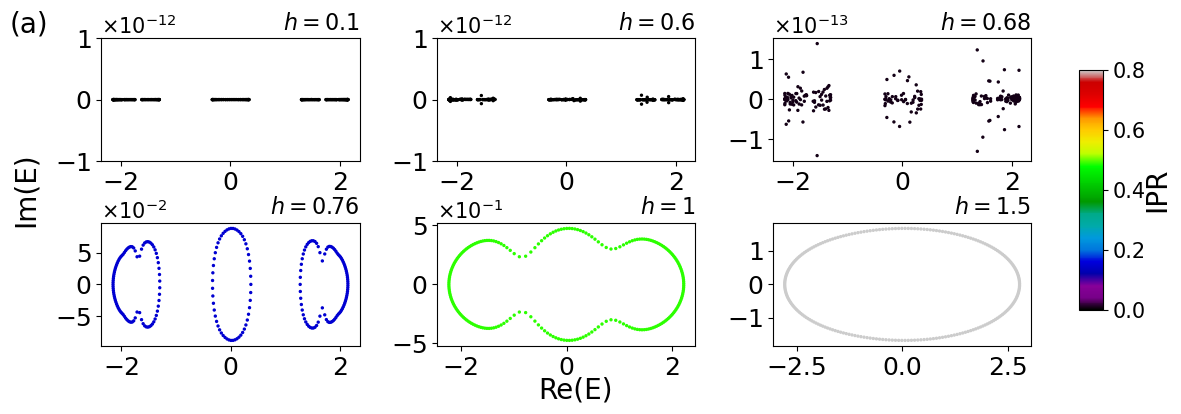}   
 \hspace*{0.1cm}
    \includegraphics[width=6cm,height=4cm]{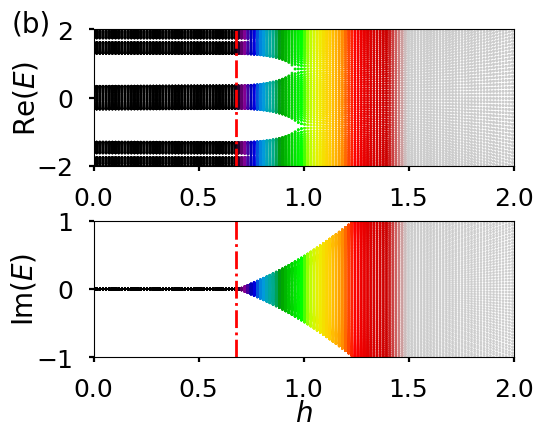}   
    \\    	
\hspace*{-12.8cm}
    \includegraphics[width=5.9cm,height=3.5cm]{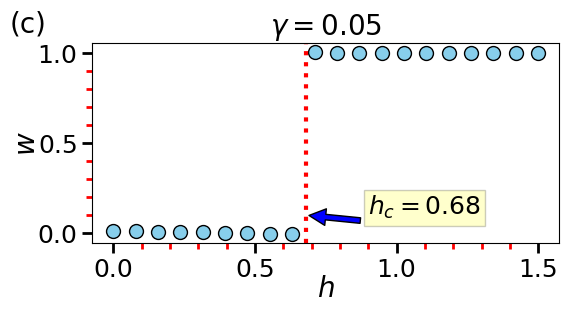}&
\hspace*{-12.2cm}
    \includegraphics[width=5.9cm,height=3.5cm]{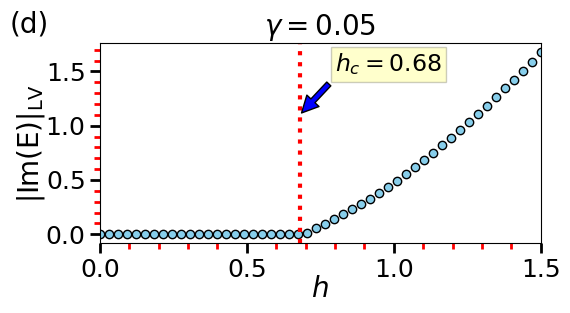}&
\hspace*{-0.3cm}
    \includegraphics[width=5.9cm,height=3.5cm]{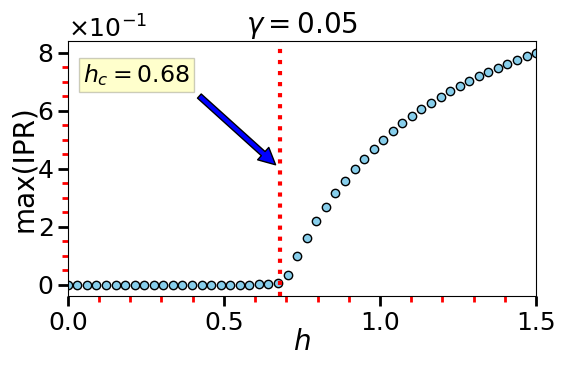}
 \\
\end{tabular}
\caption{The figure shows the eigenspectra, MI transition, PT transition, and topological transition by varying the parameter $h$, for $V=t=1.0$, and $\gamma = 0.05$. We observe three simultaneous phase transitions. Panel {\color{blue}(a)} displays the real and imaginary parts of eigenvalues (on the complex plane) and IPRs (in color scale) under the PBC for six typical values of $h$. With an increase in the value of $h$, the eigenstates tend to get localized, which is shown by color. We observe that the eigenspectra form a loop for $h>0.68$, encircling the origin of the complex energy plane, indicating a topologically non-trivial phase with non-zero windings. Also, we observe that the eigenvalues are real for $h < 0.68$, but the PT symmetry tends to break completely for $h\geq 0.68$, and thus we observe complex eigenvalues. Panel {\color{blue}(b)} shows the real parts of the eigenenergies of the Hamiltonian under PBC, where the overall energy becomes on average constant, for $h<h_c$. The bottom part of the panel displays the imaginary parts of the eigenenergies with increasing $h$. The color coding is done according to the IPR, where the red-dashed-line corresponds to the phase boundary. Panel {\color{blue}(c)} shows the winding number versus the complex phase shift $h$. The winding number makes transition transition from $0$ to $1$ as we vary $h<h_c$ to $h>h_c$. Panel {\color{blue}(d)} represents the largest value of the imaginary part of energy versus the complex phase shift $h$. For $h<h_c$, the system changes to the unbroken PT phase, where the spectrum becomes real. For $h>h_c$, there is a broken PT phase owing to the appearance of complex eigenvalues. Panel {\color{blue}(e)} represents the maximum value of IPRs contrasted to the complex phase shift $h$. For $h < h_c$, we observe the delocalized phase, marked by a zero IPR value. For $h > h_c$, all eigenstates become exponentially localized, which is marked by an increase in the IPR value. The length of the lattice is chosen to be $L=233$ for all the panels.}
\label{2} 
\end{figure*}
\end{widetext}

In Fig. \ref{2}(a), we display real and imaginary parts of the eigenvalues for different values of $h$. Here we set the parameter $\gamma = 0.05$ to observe PT symmetry transition. For $h=0.1$ and $h=0.6$, eigenvalues are real and these show PT symmetry in the system. At the critical point $h = h_c = 0.68$, we observe appearance of tiny [$\mathcal{O}(10^{-13})$] nonzero imaginary part in the eigenvalues, which indicates breaking of the PT symmetry. For $h>h_c$, multiple loops like structures begin to form in the eigenvalues spectra. In case of $h=0.76$, we observe three loops in the spectrum which display the fractal property \cite{PhysRevB.103.214202, PhysRevB.105.024514} as discussed in Appendix \ref{appendixC}. As we further increase $h=1.0,\,1.5$, we observe gradual coalescence of multiple loops and eventual formation of a single larger loop enclosing the origin. This presence of a single loop structure surrounding the origin indicates a tendency towards localization of the eigenstates. However, it does not necessarily imply full localization of all the states, as there might still be some degree of delocalization or spread in the system. As the value of $h$ increases, the eigenstates gradually become fully localized. The localization property of the eigenstates is again calculated by the IPR, and these are shown by the color coding. Moreover, this formation of loops and their coalescence with the increment of the parameter $h$ indicate an emergence of nontrivial topology in the system, which is identified by nonzero winding number. Thus Fig. \ref{2}(a) itself is indicating that the system is simultaneously making three different phase transitions or triple phase transitions. The presence of triple phase transitions will be more obvious Figs. \ref{2}(b)-(e). 

In Figure \ref{2}(b), we show the real and imaginary parts of the eigenvalues as a function of $h$, and the corresponding IPR values are shown by the color. All the eigenstates before the critical point ($h < h_c$) are extended, resulting in the robust spectrum for real and imaginary parts of the spectrum. After the critical point, the eigenvalues diverge and become complex. Here, once again, this spectral phase boundary simultaneously marks the transition of PT symmetry and MI. 
From the phase diagrams, given in Fig. \ref{Phase} and Figs. \ref{2}(a)-(b), we have got an indication that, for $\gamma = 0.05$, the system shows the triple phase transition at $h=h_c=0.68$. We are now going to show this decisively in Figs. \ref{2}(c)-(e). In Fig. \ref{2}(c), the topological  phase transition is shown by the transition in winding number $w$. The winding number is computed using Eq. \eqref{wn}. We observe that $w=0$ when $h<h_c$, and it becomes $w=1$ for $h \geq h_c$. The red dashed line shows the  transition point $h_c=0.68$. Figure \ref{2}(d) shows the PT symmetry transition, which is observed by studying the largest imaginary part of the energy eigenvalues as a function of the complex phase shift $h$. Here, we again observe PT symmetry transition at $h_c=0.68$. Finally, in Fig. \ref{2}(e), the maximum value of IPR with respect to $h$ is shown. According to our expectation, here we observe from the behavior of max(IPR) that the system makes the MI transition also at $h_c=0.68$. These results are in agreement with the phase diagram presented in Fig. \ref{Phase}. Thus we have clearly shown the triple phase transition for $\gamma = 0.05$ at $h_c=0.68$. Surprisingly, we can control the transitions and can obtain double and sequential phase transition for this model by simply varying the parameters. We have discussed this point in Appendix \ref{appendixB}. 

The prototype model, which is studied extensively in this paper, can be simulated in electric circuits using basic components like capacitors, resistors, inductors, op-amp, etc \cite{Lee2018, Bitan, Helbig2020}. In this model, the onsite potential and the hopping between neighboring sites are both non-Hermitian as well as quasi-periodic. In a recent paper, an electrical circuit is proposed to simulate non-Hermitian quasi-periodic onsite potential \cite{PhysRevResearch.2.033052}. An electric circuit design for the quasi-periodic hopping is proposed in another recent paper \cite{PhysRevB.101.020201}. Our model can similarly be simulated by combining the above mentioned electrical circuits.   

We study a non-Hermitian extension of the AAH model, which is PT symmetric. Here, the non-Hermiticity is considered at the onsite potential, as well as in the nearest-neighbor hopping term. The localization properties of the eigenstates of this model depend strongly on the boundary conditions. The system exhibits primarily localized eigenstates for the open boundary conditions. On the other hand, the MI transition occurs for the periodic boundary conditions. The disorder due to the onsite quasi-periodic potential and the non-Hermitian nature of the system contribute to these boundary-condition-sensitive behaviors. Very importantly, besides the PT symmetry and the MI transitions, we also observe the topological phase transition in this model. The overall behaviors of these phases and their transitions are shown via various phase diagrams. Interestingly, these phase diagrams indicate a parameters regime, where we may see the triple phase transition. Later, we decisively show the presence of the triple phase transition or simultaneous transitions of the three phases. Based on some recent proposals, an electric circuit based experimental simulation of this non-Hermitian model is also proposed.

\begin{acknowledgments}
JNB acknowledges financial support from DST-SERB, India through a Core Research Grant CRG/2020/001701 and also through a MATRICS grant MTR/2022/000691. Authors also thank the anonymous referee for her/his valuable comments, which helped in improving the manuscript immensely. 
\end{acknowledgments}

\appendix

\section{PT symmetry of the model}
\label{appendixA}
 
We examine the PT symmetry of the model described by the Hamiltonian \ref{1}.
The effects of the space-reflection operator $\hat{P}$ and the time-reversal operator $\hat{T}$ on a discrete system are as follows: $\hat{T}^{-1} i\, \hat{T} = -i$, and $\hat{P}^{-1} \hat{c}_j \hat{P} = \hat{c}_{L+1-j}$.
Considering the first term of the Hamiltonian.
\begin{equation}
\hat{P}^{-1}\hat{T}^{-1} V\cos(2\pi\beta j + \phi_1 + \phi_2) \hat{T}\hat{P}
\end{equation}
Substituting the value of $\phi_2 = -ih$, considering $\theta = 0$, the above equations becomes
\begin{equation}
\hat{P}^{-1}\hat{T}^{-1} V\cos(2\pi\beta j + \phi_1 - ih) \hat{T}\hat{P}.
\end{equation}
We now apply the operator $\hat{T}$ first as and then the operator $\hat{P}$, as shown above, and then simplify the above equation as follow  
\begin{equation}
\begin{split}
&\hat{P}^{-1} V\cos(2\pi\beta j + \phi_1 + ih) \hat{P} \\
&= V \cos[2\pi\beta (L+1-j) + \phi_1 + ih] \\
&= V\cos[-2\pi\beta (L+1-j) - \phi_1 - ih].
\end{split}
\end{equation}
Here, we have used the property $\cos(-\eta) = \cos(\eta)$. The onsite potential part of the Hamiltonian of the system given in Eq. \ref{1} will be PT symmetric, if the phase $\phi_1$ acquires some specific values. We calculate these values assuming that the onsite potential is PT symmetric. This means
\begin{equation}
V \cos[-2\pi\beta (L+1-j) - \phi_1 - ih] = V \cos(2\pi\beta j + \bar{\phi}_1 - ih)
\end{equation}
We find the general relation between the arguments of the cosine functions from the both sides of the above equation as
\begin{equation}
2\pi n+ [-2\pi\beta (L+1-j) - \phi_1 - ih] = 2\pi\beta j + \bar{\phi}_1 - ih
\end{equation}
and we then simplify this further as follow
\begin{equation}
\begin{aligned}
& ~2\pi n - 2\pi\beta (L+1) - \phi_1 = \bar{\phi}_1 \\
\Rightarrow & ~2\pi\beta (L+1) - (\phi_1-2\pi n) = \bar{\phi}_1 \\
\Rightarrow & ~2\pi\beta (L+1) - \phi_1\,\, ({\rm mod}\, 2\pi) = \bar{\phi}_1.
\end{aligned}
\end{equation}
In the above, we have used the following relation: if $\cos{\eta_1} = \cos {\eta_2}$, then the arguments of the cosine functions should be related in general as $2\pi n + \eta_1 = \eta_2$, where $n \in \mathbbm{Z}$. It can be shown that when $\phi_1 = \pi - \pi \beta \,\,({\rm mod}\,\, 2\pi)$ or $2\pi - \pi\beta \,\,({\rm mod}\,\, 2\pi)$, we have $\bar{\phi}_1 = \phi_1 \,\,({\rm mod}\,\, 2\pi)$. For these special values of the phase $\phi_1$, the onsite potential part of the Hamiltonian satisfies PT symmetry. Following similar steps as above and using the relation $\sin(-\eta) = - \sin(\eta)$, the SI of Ref. \cite{PhysRevA.103.L011302} has explicitly proven the PT symmetry in the hopping part of the Hamiltonian (the second and third terms of the Hamiltonian) for those special values of the phase $\phi_1$ calculated above.

\section{Double and sequential phase transitions} 
\label{appendixB}
\begin{figure}[b]
\begin{tabular}{cc}
    \includegraphics[width=0.2\textwidth]{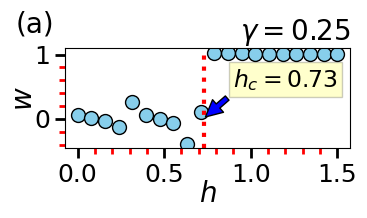} & \includegraphics[width=0.2\textwidth]{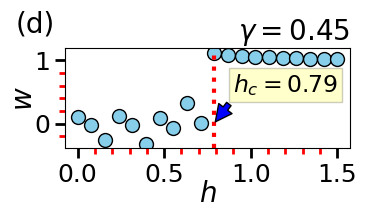} \\
    \includegraphics[width=0.2\textwidth]{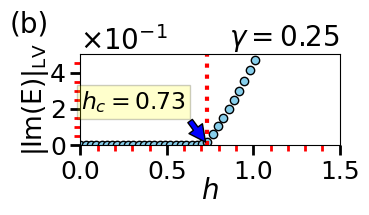} & \includegraphics[width=0.2\textwidth]{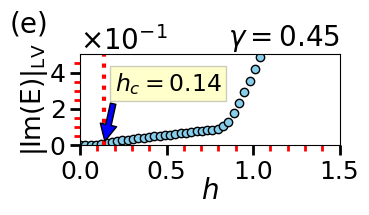} \\ 
    \includegraphics[width=0.2\textwidth]{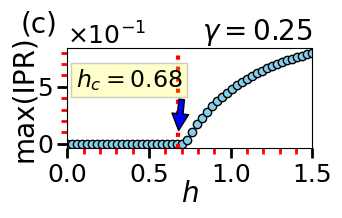} & \includegraphics[width=0.2\textwidth]{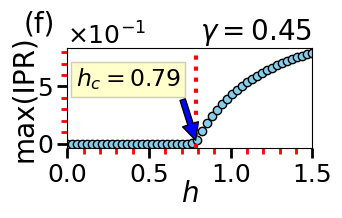} \\
\end{tabular}

\caption{Double phase transition. Here we set for all the panels the number of lattice sites $L=233$, and the parameters $V = t = 1.0$. The rational approximation of the irrational number is chosen as $\beta = \frac{144}{233}$. (a)-(b): the PT symmetry and the topological phase transition occur simultaneously at $h_c=0.73$, for the parameter $\gamma = 0.25$. (c) However, the MI transition occurs at $h_c=0.68$. (d) and (f): The topological and the MI transition occur simultaneously at $h_c=0.79$, when $\gamma = 0.45$. (e) The PT symmetry transition occurs at $h_c=0.14$.}
\label{A1} 
\end{figure}

Here, we focus on the study of the variation of the system parameters that lead to different or similar transition points. The case of the triple phase transition of the above Hamiltonian is already studied in the main text. We now focus on the cases, where we observe double and sequential phase transitions. 
During a careful investigation the results presented in Fig. 2 of the main text, we also chose two other values of the non-Hermitian parameter $\gamma = 0.25$ and $0.45$. For the case of $\gamma = 0.25$, we observe simultaneous transition of the PT symmetry and the topological at $h_c = 0.73$, while the MI transition or localization-delocalization transition occurs at the lower value $h_c=0.68$. However, for the other case when $\gamma = 0.45$, the PT transition occurs at $h_c = 0.14$, while the MI and the topological transitions are observed at $h_c = 0.79$. This different types of double phase transitions are shown in Fig. \ref{A1}

\begin{figure}[t]
\begin{tabular}{cc}
    \includegraphics[width=0.2\textwidth]{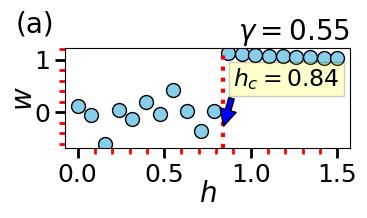} &     \includegraphics[width=0.2\textwidth]{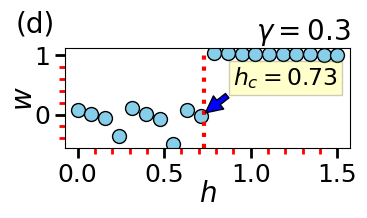} \\
    
    \includegraphics[width=0.2\textwidth]{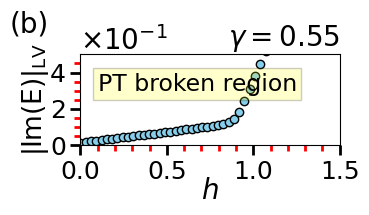} & \includegraphics[width=0.2\textwidth]{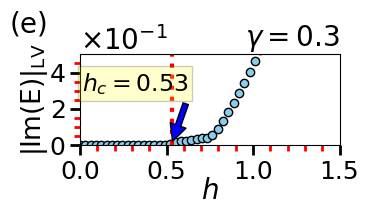}\\
    \includegraphics[width=0.2\textwidth]{{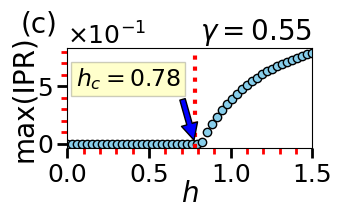}}&
     \includegraphics[width=0.2\textwidth]{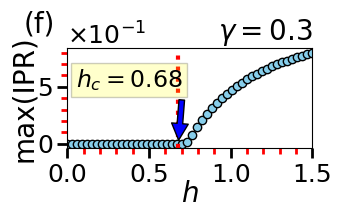}

\end{tabular}

\caption{Sequential phase transition. Here again we set $L=233$, $V = t = 1.0$ and $\beta = \frac{144}{233}$ for all the panels. (a)-(c): PT symmetry broken region. Topological transition occurs at $h_c = 0.84$, while the MI transition occurs at $h_c=0.78$. (d)-(f): PT symmetric region. Here, the PT transition is observed at $h_c=0.53$ (e), followed by the MI transition at $h_c=0.68$ (d), and finally the topological phase transition occurs at $h_c=0.73$ (f).}
\label{A2} 
\end{figure}

In order to observe the sequential transition, we select the value of the parameter $\gamma$ in both PT broken region, as well as in the PT unbroken region. For example, when $\gamma=0.55$, the Hamiltonian in Eq. (\ref{1}) is not a PT symmetric for any value of the other non-Hermitian parameter $h$. However, we observe the topological and the MI transition at different critical points. On the other hand, for $\gamma=0.3$, the Hamiltonian is PT symmetric. For this case, we observe transitions of the three phases at three different values of $h_c$. Based on the above observation, we realize that PT transition precedes, while the topological transition is posterior to all the three transitions as shown in Fig. \ref{A2}. 

\section{Illustration of multi-fractal, extended (or delocalized), and localized behavior} 
\label{appendixC}
\begin{figure}[b]
  \centering
  \begin{tabular}{cc}
    \includegraphics[width=4cm,height=4cm]{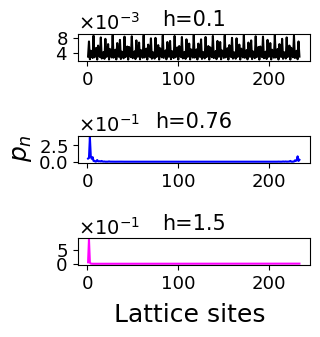} &
    \includegraphics[width=4cm,height=4cm]{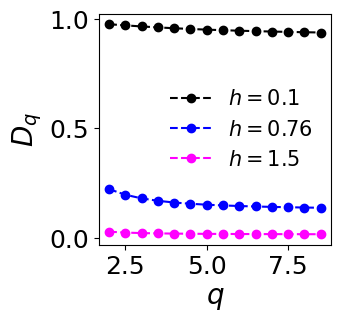} \\
      ~~~~~~~~(a) &  ~~~~~~~~(b)
  \end{tabular}
  \caption{Subfigure (a) shows the occupation number as a function of lattice site for three different values of the non-Hermitian parameter, $h=0.1$, $h=0.76$, and $h=1.5$. (b) shows the generalized fractal dimensions $D_q$ for $h=0.1$, $h=0.76$, and $h=1.5$. The plot for $h=0.1$ indicates extended behavior, with $D_q = 1$ for increasing $q$. For $h=0.76$, the plot shows multifractal behavior with a decrease in $D_q$ as $q$ increases. Finally, the plot for $h=1.5$ indicates localized behavior with $D_q=0$ for increasing values of $q$.}
  \label{A3}
\end{figure}
In the main text, we have observed that the eigen-spectra for the parameter value $h=0.76$ display fractal behavior with two loops, which are not encircling the origin. Another loop, which encircles the origin, the corresponding eigenstates show localized behavior. Here we elaborate that observation by presenting additional results for the onsite occupation number $p_{n,j} = |u_{n,j}|^2$ and the generalized fractal dimensions $D_q$ \cite{10.21468}. Here, $n$ denotes the quantum number associated with the energy level and $j$ denotes the lattice site index. We generalize the notion of IPR by considering the following relation 
\begin{equation}
\sum_{j=1}^{L}(p_{0,j})^q \sim L^{-\tau_q}, \label{eq:2}
\end{equation}
where $\tau_q=D_q(q-1)$. Here, the moment $q$ is only restricted to positive integer values with $q \geq 2$ and the $q$-fractal dimension $D_q$ is calculated to see the multi-fractality in the spectrum \cite{PhysRevB.106.024204}. For the delocalized and localized states, $D_q$ will be independent of $q$. For the former case $D_q=1$, while for the latter case $D_q=0$.

Figure \ref{A3}(a) shows the occupation number as a function of lattice site for three different values of the non-Hermitian parameter $h=0.1$, $h=0.76$, and $h=1.5$. The result of $h=0.1$ case indicates extended or delocalized behavior of the state; while for $h=0.76$ and $h=1.5$, the spectrum respectively shows multi-fractal (showing more than one peaks) and localized behavior (showing only one peak). Furthermore, Fig. \ref{A3}(b) presents the generalized fractal dimensions $D_q$ as a function of the order $q$ for the same three values of $h$. The case of $h=0.1$ exhibits constant fractal dimension $D_q = 1$ with increasing $q$. This indicates extended behavior of the eigenstates. In contrast, the plot for $h=0.76$ displays a decrease in the value of $D_q$ with an increase in $q$, consistent with multi-fractal behavior. For the parameter value $h=1.5$, $D_q=0$ with increasing values of $q$, which indicates localized behavior of the state.

\section{Hofstadter butterfly like fractal spectrum} 
\label{appendixD}
\begin{figure}[t]
\centering
\begin{tabular}{ccc}

\hspace{-0.5cm}
  \includegraphics[width=0.21\textwidth]{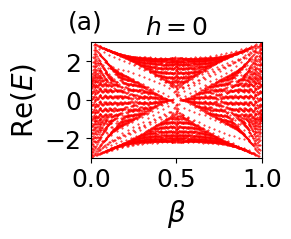}\hspace{-0.45cm}
	
    \includegraphics[width=0.17\textwidth]{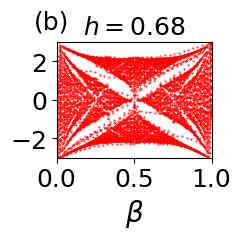}\hspace{-0.45cm}

   		 \includegraphics[width=0.17\textwidth]{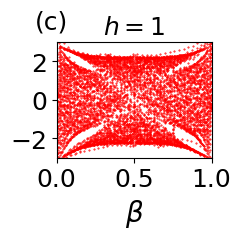}\\[-3ex] 
 
\end{tabular}
\caption{The band spectrum of the Hamiltonian \ref{1} shows a fractal behavior, called Hofstadter’s butterfly, under open boundary condition for, $L=34$, $t$, $V=1$ and $\gamma = 0.05$. The width of the spectrum depends on the value of $t$ and $V$. We can see that the spectral nature disappears while we increase the non-Hermiticity of the system.}
\label{A4}   
\end{figure}

One of the important and very generic properties of the spectrum of different versions of the Aubry-Andr\'e (AA) model is its fractal nature \cite{yuce2014pt,martinez2018quasiperiodic}. It is observed that a similar fractal nature for the two dimensional generalized non-Hermitian AA model also exist. On further investigation, we deduce the spectrum as a function of the parameter $\beta$, and observe a Hofstadter butterfly like structure \cite{hofstadter1976energy}. 
Since, the system repeats itself in equal intervals of $\beta$, therefore it is adequate enough to analyze the region $0<\beta<1$. Furthermore, the energy spectrum is symmetric with respect to $\beta = 0.5$ axis as shown Fig. \ref{A4}. We observe that as we increase the non-Hermiticity in the sysyem, the fractal nature of the spectrum tends to disappear.

%

\end{document}